\documentclass[a4paper,11pt]{article}%
\usepackage{amssymb,amsmath,amsfonts}
\usepackage{graphicx}
\usepackage{amsmath}
\usepackage{amsfonts}
\usepackage{amssymb}%
\providecommand{\U}[1]{\protect\rule{.1in}{.1in}}
\newtheorem{theorem}{Theorem}

\newtheorem{corollary}[theorem]{Corollary}

\newtheorem{idea memo}[theorem]{Idea Memo}

\newtheorem{remark}[theorem]{Remark}

\begin{document}

\title{Who has seen a free photon?}
\author{Izumi Ojima \thanks{E-mail: ojima@kurims.kyoto-u.ac.jp}\\and \\Hayato Saigo\thanks{E-mail: harmonia@kurims.kyoto-u.ac.jp } \\Research Institute for Mathematical Sciences, \\Kyoto University, Kyoto 606-8502, Japan}
\date{}
\maketitle

\begin{abstract}
While the notion of the position of photons is indispensable in the quantum
optical situations, it has been known in mathematical physics that any
position operator cannot be defined for a massless free particle with a
non-zero finite spin. This dilemma is resolved by introducing the
\textquotedblleft effective mass\textquotedblright\ of a photon due to the
interaction with matter. The validity of this interpretation is confirmed in
reference to the picture of \textquotedblleft polariton\textquotedblright, a
basic notion in optical and solid physics. In this connection, we discuss the
long-standing controversy between Minkowski's and Abraham's definitions of the
momenta of a photon in media from the general viewpoint adopted in the
appendix. \quad

\end{abstract}

\section{A dilemma about localizability of photons}

In the recent advanced quantum-optical technology, the notion of positions of
photons has played indispensable roles in the experimental situations, as is
exemplified by the most familiar Mach-Zender interferometry consisting of the
beam splitters and of the photon detector to specify the position of detection
of photons. Theoretically speaking, this is the contexts where the
localization of a photon is to be described by means of some \textit{position
observables} associated with the quantum electromagnetic field. Here we
encounter the following serious difficulty: since the paper of Newton and
Wigner \cite{N-W}, it has been known in mathematical physics that any position
operator cannot be defined for a massless free particle with a non-zero finite
spin, in sharp contrast to the cases of massive particles which can be
localized. This statement is clearly in contradiction to the above familiar
situations where almost all physicists have used the notion of
\textquotedblleft position of a photon\textquotedblright\ as one of the basic
ingredients of theory and application of quantum mechanics. Then, who has seen
a free photon?

Once Dirac wrote in \cite{DIR} as follows: \vskip8pt

\begin{quote}
If we are given a beam of roughly monochromatic light, then we know something
about the location and momentum of the associated photons. We know that each
of them is located somewhere in the region of space through which the beam is
passing ...
\end{quote}

\vskip8pt This modest statement contradicts, however, the above general
result, if it is literally interpreted. Namely, if we try to make sense out of
the sentence, \textquotedblleft each free photon is located somewhere in the
region of space\textquotedblright, we need to introduce a \textquotedblleft
position operator for a free photon in three-dimensional space as a well
defined observable\textquotedblright, which is in contradiction with
\cite{N-W}. In the present paper, we propose a solution to the problem of
localization of a photon taking proper account of the relevant physical
interactions between a photon to be localized and the matter to localize the
former, combining the basic ideas relevant to both fields.

\section{Strategy to solve the dilemma}

We propose the following resolution: while a massless free photon is
\textit{not} localizable, a \textquotedblleft real\textquotedblright\ photon
can be made localizable by its dynamical interaction with matter such as media
or devices (for its detection). To be precise, this simple picture can be
shown to have a sound basis as follows. First we see in Section 3 that a
photon can be localized only if photon-matter interaction provides a positive
effective mass, by our reinterpretation of Wightman's theorem \cite{WIG}
following from the arguments by Newton and Wigner \cite{N-W}. Then we show in
Section 4 how the photon-matter interaction can be reduced in this context
into the picture of \textquotedblleft a free particle with positive effective
mass\textquotedblright. This interpretation becomes clearer if we refer to the
picture of \textquotedblleft polariton\textquotedblright\cite{FAN,HOP}, a
basic notion in optical and solid-state physics.

This kind of new connection between mathematical physics and other region of
physics should bring fruitful perspectives. We conclude this paper with a
re-interpretation of Dirac's statement above and further suggestions.

\section{Formulation of the problem}

\subsection{Newton-Wigner-Wightman analysis}

In 1949, Newton and Wigner \cite{N-W} raised the question of localizability of
single free particles. They attempted to formulate the properties of the
localized states on the basis of natural requirements of relativistic covariance.

\paragraph{}

Physical quantities available in this formulation admitting direct physical
meaning are restricted inevitably to the generators of Poincar\'{e} group
$\mathcal{P}_{+}^{\uparrow}=\mathbb{R}^{4}\rtimes L_{+}^{\uparrow}$ (with
$L_{+}^{\uparrow}$ the orthochronous proper Lorentz group) which is locally
isomorphic to the semi-direct product $\mathcal{H}_{2}(\mathbb{C})\rtimes
SL(2,\mathbb{C})$ of the hermitian (2$\times$2)-matrices and of
$SL(2,\mathbb{C})$, consisting of the energy-momentum vector $P_{\mu}$ and of
the Lorentz generators $M_{\mu\nu}$ (composed of angular momenta $M_{ij}$ and
of Lorentz boosts $M_{0i}$). The problem is then to find conditions under
which \textquotedblleft position operators\textquotedblright\ can naturally be
constructed from the Poincar\'{e} generators $(P_{\mu},M_{\mu\nu})$. In
\cite{N-W}, position operators have been shown to exist in massive cases in an
essentially unique way for \textquotedblleft elementary\textquotedblright%
\ systems in the sense of the irreducibility of the corresponding
representations of $\mathcal{P}_{+}^{\uparrow}$ so that localizability of a
state can be defined in terms of such position operators. In massless cases,
however, no localized states are found to exist in the above sense. That was
the beginning of the story.

Wightman \cite{WIG} clarified the situation by recapturing the notion of
\textquotedblleft localization\textquotedblright\ in a general form as
follows. First he has reformulated the usual approaches in terms of unbounded
position operators into the form of general axioms (i)-(v) involving
projection operators,

\begin{description}
\item[(i)] To each Borel subset $\Delta$ of $\mathbb{R}^{3}$, there
corresponds a projection operator $E(\Delta)$ in a Hilbert space
$\mathfrak{H}$, whose expectation value gives the probability of finding the
system in $\Delta$;

\item[(ii)] $E(\Delta_{1}\cap\Delta_{2})=E(\Delta_{1})E(\Delta_{2})$;

\item[(iii)] $E(\Delta_{1}\cup\Delta_{2})=E(\Delta_{1})+E(\Delta_{2})$, if
$\Delta_{1}\cap\Delta_{2}=\phi$;

\item[(iv)] $E(\mathbb{R}^{3})=1$;

\item[(v)] $E(\mathcal{R}\Delta+\mathbf{a})=U(\mathbf{a},\mathcal{R}%
)E(\Delta)U(\mathbf{a},\mathcal{R})^{-1}$, where $\mathcal{R}\Delta
+\mathbf{a}$ is the set obtained from $\Delta$ by applying a translation
$\mathbf{a}$ after a rotation $\mathcal{R}$, and $U(\mathbf{a},\mathcal{R})$
is the corresponding unitary operator in $\mathfrak{H}$.
\end{description}

Note that the notion of localizability discussed above is concerned with
\textit{localization of states in space at a given time}. If we consider the
axioms like (i)-(v) on the whole space-time, it would imply the validity of
the CCR relations between 4-momenta $p_{\mu}$ and space-time coordinates
$x^{\nu}$, which would imply the Lebesgue spectrum covering the whole
$\mathbb{R}^{4}$ for both observables $\hat{p}_{\mu}$ and $\hat{x}^{\nu}$.
Then, any such physical requirements as the spectrum condition cannot be
imposed on the energy-momentum spectrum $\hat{p}_{\mu}$, and hence, the notion
of localizability in space-time does not make sense.

According to Mackey's theory of induced representations, Wightman's
formulation can easily be seen as the condition for the set of operators
$\left\{  E(\Delta)\right\}  $ to constitute a \textit{system of
imprimitivity} \cite{MAC} under the action of the unitary representation
$U(a,\mathcal{R})$ in $\mathfrak{H}$ of the three-dimensional Euclidean group
$SE(3):=\mathbb{R}^{3}\rtimes SO(3)$. In a more algebraic form, the pair
$(E,U)$ can also be viewed as a \textit{covariant representation}
\begin{equation}
E(\tau_{(\mathbf{a},\mathcal{R)}}(f))=U(a,\mathcal{R})E(f)U(a,\mathcal{R}%
)^{-1}\text{ \ \ \ for }f\in L^{\infty}(\mathbb{R}^{3}),(\mathbf{a}%
,\mathcal{R)\in}SE(3), \label{imprim}%
\end{equation}
of an action $SE(3)\underset{\tau}{\curvearrowright}L^{\infty}(\mathbb{R}%
^{3})$, $[\tau_{(\mathbf{a},\mathcal{R)}}(f)](\mathbf{x}):=f(\mathcal{R}%
^{-1}(\mathbf{x}-\mathbf{a}))$ on the algebra $L^{\infty}(\mathbb{R}^{3})$
generated by the position operators in the representation $E:L^{\infty
}(\mathbb{R}^{3})\ni f\longmapsto E(f)=\int f(\mathbf{x})dE(\mathbf{x})\in
B(\mathfrak{H})$, s.t. $E(\chi_{\Delta})=E(\Delta)$.

Thus Wightman's formulation of the Newton-Wigner localizability problem is
just to examine whether the Hilbert space $\mathfrak{H}$ of the representation
$(U,\mathfrak{H})$ of $SE(3)$ can accommodate a representation $E$ of the
algebra $L^{\infty}(\mathbb{R}^{3})$ consisting of position operators,
covariant under the action of $SE(3)$ in the sense of (\ref{imprim}) .

Applying the general theory of Mackey to the case of three-dimensional
Euclidean group $SE(3)$, Wightman proved the fundamental result below as a
purely kinematical consequence.

\begin{theorem}
[\cite{WIG}, excerpt from theorem 6 and 7]A Lorentz or Galilei covariant
massive system is always localizable. For the Lorentz case, the only
localizable massless elementary system (i.e. irreducible representation) has
spin zero. For the Galilei case, no massless elementary system is localizable.
\end{theorem}

\begin{corollary}
A free photon is not localizable.
\end{corollary}

The essential mechanism of (non-)localizability in the sense of
Newton-Wigner-Wightman depends\ on the structure of little groups defined by
Wigner as the stabilizer groups of standard four-momenta on each type of
$\mathcal{P}_{+}^{\uparrow}$-orbits in $p$-space.

On an orbit $p^{2}=m^{2}c^{2}>0$ under $\mathcal{P}_{+}^{\uparrow}$, we can
choose a standard momentum $p^{(0)}:=(mc,\mathbf{0})$ which specifies a rest
frame of a particle with mass $m\neq0$. Then, the little group at $p^{(0)}$ is
the group $SO(3)$ of spatial rotations, corresponding to the degrees of
freedom remaining
in the rest frame. As a consequence, \textquotedblleft the space of all
Lorentz frames\textquotedblright\ along the orbit becomes $SO(1,3)/SO(3)\cong%
\mathbb{R}^{3}$. Note that a Lorentz boost $\Lambda_{p}$ defined by
\[
\Lambda_{p}=\left(
\begin{array}
[c]{cc}%
\dfrac{p^{0}}{mc} & \dfrac{^{t}\mathbf{p}}{mc}\\
\dfrac{\mathbf{p}}{mc} & \mathbf{1}+(\dfrac{p^{0}}{mc}-1)\dfrac{\mathbf{p}%
^{t}\mathbf{p}}{\mathbf{p}^{2}}%
\end{array}
\right)  \in SO(1,3)
\]
transforms $p^{(0)}$ into $p=\Lambda_{p}p^{(0)}=(p^{0},\mathbf{p}%
)=(\sqrt{m^{2}c^{2}+\mathbf{p}^{2}},\mathbf{p})$. On the other hand, we have
for $p_{\Lambda}=\Lambda p^{(0)}$ a relative velocity $\mathbf{u}_{\Lambda}%
\in\mathbb{R}^{3}\cong SO(1,3)/SO(3)$ between the Lorentz frames
$(1,\mathbf{0})$ and $\dfrac{p_{\Lambda}}{mc}=:u_{\Lambda}=(\dfrac{1}%
{\sqrt{1-\mathbf{u}_{\Lambda}^{2}/c^{2}}},\dfrac{\mathbf{u}_{\Lambda}/c}%
{\sqrt{1-\mathbf{u}_{\Lambda}^{2}/c^{2}}}),\\ u_{\Lambda}^{2}=1$. Thus, the
homeomorphism $\mathbb{R}^{3}\cong SO(1,3)/SO(3)$ describes a non-trivial
action of $\mathbf{u}=\mathbf{u}_{\Lambda}\in\mathbb{R}^{3}$ on $p=(p^{0}%
,\mathbf{p})$ belonging to the $\mathcal{P}_{+}^{\uparrow}$-orbit $p^{2}%
=m^{2}c^{2}>0$ through the action of Lorentz boost $\Lambda=\Lambda
(\mathbf{u})=\Lambda_{p}\in SO(1,3)$ transforming $p^{(0)}$ into $p=\Lambda
p^{(0)}$. Hence the coordinates $\mathbf{u}\in\mathbb{R}^{3}\cong
SO(1,3)/SO(3)$ of Lorentz frames\ just play the role of the order parameters
(or, \textquotedblleft sector parameters\textquotedblright) on each
$\mathcal{P}_{+}^{\uparrow}$-orbit as the space of \textquotedblleft
condensation\textquotedblright\ associated with a symmetry breaking of boost
invariance, and hence, $\mathbf{u}$ can be identified with position operators
in the imprimitivity system appearing in Wightman's theorem. 

In sharp contrast, there is no rest frames for a massless particle and the
little group becomes isomorphic to two-dimensional Euclidean group $SE(2)$,
whose rotational generator corresponds to the helicity. Since the other two
translation generators corresponding to gauge transformations span
\textit{non-compact} directions in distinction from the massive cases with
compact $SO(3)$, the allowed representation is only the trivial one which
leaves the transverse modes invariant, and hence, the little group cannot
provide position operators in the massless case.

Since Newton-Wigner-Wightman, many discussions around the photon localization
problem have been developed. So far as we know, the arguments seem to be
divided into two opposite viewpoints, one relying on purely dynamical bases
\cite{HAA} and another on pure kinematics \cite{FLA}, where it is almost
impossible to find any meaningful agreements. Below we propose an alternative
strategy based on the notion of \textquotedblleft effective
mass\textquotedblright, which can provide a reasonable reconciliation between
these conflicting ideas because of its \textquotedblleft
kinematical\textquotedblright\ nature arising from some dynamical origin.

\subsection{Wightman's theorem as the \textquotedblleft
basis\textquotedblright\ for localization}

Our scheme of the localization for photons can be summarized as follows, which
is essentially in accordance with the basic formulation of \textquotedblleft
quadrality scheme\textquotedblright\ \cite{OJI2} underlying the Micro-Macro
duality \cite{OJI1}:
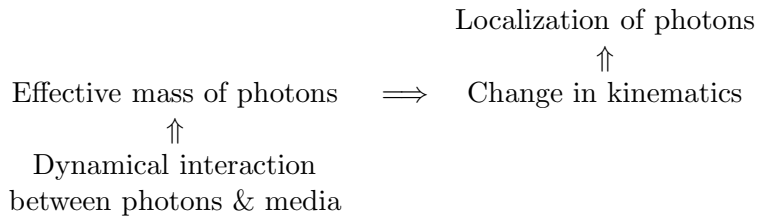
\[%
\begin{array}
[c]{ccc}
&  & \text{Localization of photons}\\
&  & \Uparrow\\
\text{Effective mass of photons} & \Longrightarrow & \text{Change in
kinematics}\\
\Uparrow &  & \\%
\begin{array}
[c]{c}%
\text{Dynamical interaction}\\
\text{between photons \& media}%
\end{array}
&  &
\end{array}
\]
\newline Once a positive effective mass appears, Wightman's theorem itself
provides the \textquotedblleft kinematical basis\textquotedblright\ for the
localization of a photon. From our point of view, therefore, this theorem so
far regarded as a no-go theorem against the localizability becomes actually an
affirmative support for it, conveying such a strongly selective meaning that,
whenever a photon is localized, it should carry\ a non-zero effective mass.

In the next section, we explain the meaning of our scheme from a physical
point of view.

\section{Resolution of the dilemma}

\subsection{How to define effective mass of a photon}

Now we focus on a photon interacting with homogeneous medium, in the case of
the monochromatic light with angular frequency $\omega$ as a classical light
wave. For simplicity, we neglect here the effect of absorption, that is, the
imaginary part of refractive index. When a photon interacting with matter can
be treated as a single particle, it is natural to identify its velocity
$\mathbf{v}$ with the \textquotedblleft signal velocity\textquotedblright\ of
light in medium. The relativistic total energy of the particle $E$ should be
related to $v:=\sqrt{\mathbf{v\cdot v}}$ by its mass $m_{\mathsf{eff}}$:
\begin{equation}
E=\frac{m_{\mathsf{eff}}c^{2}}{\sqrt{1-\frac{v^{2}}{c^{2}}}} \label{energy}%
\end{equation}
Since $v$ is well known to be smaller than the light velocity $c$
(theoretically or experimentally), $m_{\mathsf{eff}}$ is positive (when the
particle picture above is valid). Then we may consider $m_{\mathsf{eff}}$ as
the relativistic \textquotedblleft effective (rest) mass of a
photon\textquotedblright, and identify its momentum $\mathbf{p}$ with
\begin{equation}
\text{$\mathbf{p}$}=\frac{m_{\mathsf{eff}}\text{$\mathbf{v}$}}{\sqrt
{1-\frac{v^{2}}{c^{2}}}}. \label{momentum1}%
\end{equation}
Hence, as long as \textquotedblleft an interacting photon\textquotedblright%
\ can be approximately treated as a single particle, it should be massive,
according to which its \textquotedblleft localization
problem\textquotedblright\ is resolved. The validity of this picture will be
confirmed in the next subsection.

The concrete forms of energy/momentum are related to the Abraham-Minkowski
controversy \cite{ABR, MIN, BAR} and modified versions of Einstein/de Broglie
formulae. (We discuss this point in the appendix.)

Our argument itself, however, does not depend on the energy/momentum formulae.
The only essential point is that the interactoin can make a massive particle
from a massless one. That is, while a free photon satisfies
\begin{equation}
E_{\mathrm{free}}^{2}-c^{2}p_{\mathrm{free}}^{2}=0,
\end{equation}
an interacting photon satisfies
\begin{equation}
E^{2}-c^{2}p^{2}=m_{\mathsf{eff}}^{2}c^{4}>0.
\end{equation}
To sum up, an \textquotedblleft interacting photon\textquotedblright\ can gain
a positive effective mass, while a \textquotedblleft free
photon\textquotedblright\ remains massless! This is the key we have sought
for. However, the argument in this section is based on the assumption that
\textquotedblleft a photon dressed with interaction\textquotedblright\ can be
viewed as a single particle.
Then we proceed to consider the validity of our picture, especially the
existence of particles whose effective mass is obtained by the interaction,
analogous to Higgs mechanism: Such a universal model for photon localization
exists. It is the notion of polariton, well known in optical and solid physics.

\subsection{Polaritons as a universal model for photon localization}

In optical and solid-state physics, the propagation of light in a medium is
viewed as follows: By the interaction between light and matter, creation of an
\textquotedblleft exciton (an excited state of polarization field above the
Fermi surface)\textquotedblright\ and annihilation of a photon will be
followed by annihilation of an exciton and creation of a photon, ..., and so
on. This chain of processes itself is often considered as the motion of
particles called polaritons (in this case \textquotedblleft
exciton-polaritons\textquotedblright), which constitute particles associated
with the coupled wave of the polarization wave and electromagnetic wave.

\begin{remark}
In spite of the similarity in its name, a polariton should not be confused
with a polaron \cite{LAN} which represents a fermion as a charged matter
dressed by polarization field. In contrast, a polariton is a boson which
represents a \textquotedblleft dressed photon\textquotedblright.
\end{remark}

The notion of polariton has been introduced to develop the microscopic theory
of electromagnetic interactions in materials (\cite{FAN}, \cite{HOP}). An
injected photons become polaritons by the interaction with matter. As
exiton-phonon interaction is dissipative, the polariton picture gives a
scenario of absorption. It has provided a better approximation than the
scenarios without a polariton. Moreover, the group velocity of polaritons
discussed below gives another confirmation of the presence of an effective mass.

As is well known, permittivity $\epsilon(\omega)$ is given by the following
equality,%
\begin{equation}
\epsilon(\omega)=n^{2}=\frac{c^{2}k^{2}}{\omega^{2}},
\end{equation}
and hence, we obtain the dispersion relation (a relation between frequency and
wave number) of polariton once the formula of permittivity is given.

\begin{remark}
In general, this dispersion relation implies branching, analogous to the Higgs
mechanism. The signal pulse correponding to each branches can also be detected
in many experiments, for example, in \cite{MAS} cited below.
\end{remark}

In the simple case, the permittivity is given by the transverse frequency
$\omega_{T}$ of exciton's (lattice vibration) as follows:
\begin{equation}
\epsilon(\omega)=\epsilon_{\infty}+\frac{\omega_{T}^{2}(\epsilon_{st}%
-\epsilon_{\infty})}{\omega_{T}^{2}-\omega^{2}},
\end{equation}
where $\epsilon_{\infty}$ denotes $\lim_{\omega\rightarrow\infty}%
\epsilon(\omega)$ and $\epsilon_{st}=\epsilon(0)$ (static permittivity). With
a slight improvement through the wavenumber dependence of the exciton energy,
the theoretical result of polariton group velocity $\frac{\partial\omega
}{\partial\mathbf{k}}<c$ based on the above dispersion relation can explain
satisfactorily experimental data of the passing time of light in materials
(for example, \cite{MAS}). This strongly supports the validity of the
polariton picture.

From the above arguments, polaritons can be considered as a universal model of
the \textquotedblleft interacting photons in a medium\textquotedblright\ in
the previous subsection 4.1. The positive mass of a polariton gives a solution
to its \textquotedblleft localization problem\textquotedblright. Conversely,
as the \textquotedblleft consequence\textquotedblright\ of Wightman's theorem,
it follows that \textquotedblleft all\textquotedblright\ physically accessible
photons as particles which can be localized are more or less polaritons (or
similar particles) because only the interaction can give a photon its
effective mass, if it does not violate particle picture. In this way, the
dilemma between Newton-Wigner-Wightman theorem and the position observable of
photons is successfully resolved by combining useful mathematical methods and
meaningful physical concepts, which were separated before causing a negative result.

\section{Concluding remarks}

Now the statement by Dirac in the beginning section can be justified in the
following form: If we are given a beam of roughly monochromatic light, then we
know something about the location and momentum of the associated photons with
\textquotedblleft effective mass\textquotedblright\ (polaritons) arising from
their interactions with matter media. We know that each of them is located
somewhere in the region of space \textquotedblleft filled with a medium (like
the air or crystals)\textquotedblright\ through which the beam is passing ...

This modified statement and all the discussions in the present paper clarify
the important roles played by interactions in making sense of the notion of
localization. We can expect that the discussion in the present paper will shed
some new light on the idea of the \textquotedblleft
emergence\textquotedblright\ of space-time proposed in \cite{OJI2}. In
combination with a possible scenario for the mass generation, we can summarize
the argument above in the following quadrality scheme:
\[%
\begin{array}
[c]{ccc}
&  & x:%
\begin{array}
[c]{c}%
\text{localization }\\
\text{of photons}%
\end{array}
\\
&  & \Uparrow\\
\text{ }m_{\mathrm{eff}}:\text{Effective mass of photon} & \Longrightarrow &
v:\text{kinematics }\\
\Uparrow &  & \\
p:%
\begin{array}
[c]{c}%
\text{dynamical interaction}\\
\text{between photons \& media}%
\end{array}
&  &
\end{array}
\]
When all the above ingredients are established, the \textquotedblleft
mechanics of mass points\textquotedblright\ becomes meaningful. A photon,
which is something quite dissimilar to a \textquotedblleft mass
point\textquotedblright, appears ubiquitously since the electromagnetic field
works as a universal medium to mediate the interaction between charged
particles which provides an idealized standard reference system. The answer to
our question at the beginning can now be found in the following modified form
of the famous verse \cite{WIND}:

\begin{quote}
Who has seen a free photon?

Neither I nor you.

But when the matter reacts trembling

the photons are passing through.
\end{quote}

\section*{Acknowledgements}

The authors are grateful to Prof. Shogo Tanimura, who has informed them of the
importance of the dilemma. They thank Mr. Ryo Harada for many useful comments
to refine drafts. They also thank Mr. Hiroshi Ando, Mr. Takahiro Hasebe, and
Mr. Kazuya Okamura for their interests and comments.

\section*{Appendix: A new viewpoint on Minkowski-Abraham controversy and
generalized Einstein-de Broglie formula}

We consider in this appendix the problem as to how to express the effective
mass $m_{\mathsf{eff}}$ of a photon (a polariton) in a medium precisely, which
can be easily answered if $E$ or $\mathbf{p}$ is identified. At this point,
however, we encounter another mystery which has been the origin of a
long-standing controversy for one century(!), known as \textquotedblleft
Abraham-Minkowski dilemma\textquotedblright\ \cite{BAR} concerning the correct
formula for the momentum of a photon in medium: in 1908 Minkowski \cite{MIN}
proposed a candidate for the energy momentum tensor of the electromagnetic
field in a medium, according to which the photon momentum takes such a form
as
\begin{equation}
p=p_{\mathrm{free}}\cdot n,
\end{equation}
where $p:=\sqrt{\mathbf{p}\cdot\mathbf{p}}$ is the magnitude of the momentum
and $p_{\mathrm{free}}$ the free photon momentum. $n$ denotes the refractive
index and $v_{\mathsf{ph}}$ is the magnitude of phase velocity of light in the
medium, as is given as usual by
\begin{equation}
v_{\mathsf{ph}}=\frac{\omega}{k}=\frac{c}{n},
\end{equation}
where $k$ and $\omega$ are, respectively, the magnitude of the wavenumber
$\mathbf{k}$ and the frequency of the (classical) light wave in the medium.
When $v_{\mathsf{ph}}<c$ or equivalently $n>1$ (as it should be in the normal
situations), $p=p_{Min}$ is larger than the free photon momentum
$p_{\mathrm{free}}$. On the other hand, Abraham \cite{ABR} proposed in 1909
another version leading to a formula for the momentum $p=p_{Abr}$ of a photon
in a medium given by%
\begin{equation}
p=p_{\mathrm{free}}\cdot\frac{v}{c},
\end{equation}
which is smaller than pure photon momentum.

To settle the matter, we start from the Minkowski-type momentum for a photon
in a medium:
\begin{equation}
\text{$\mathbf{p}$}=\hbar\text{$\mathbf{k.}$} \label{momentum2}%
\end{equation}
with
\begin{equation}
p=p_{\mathrm{free}}\cdot n=p_{\mathrm{free}}\cdot\frac{c}{v_{\mathsf{ph}}%
}=\frac{\hbar\omega}{v_{\mathsf{ph}}}.
\end{equation}
Then the effective mass $m_{\mathsf{eff}}$ is given by
\begin{equation}
m_{\mathsf{eff}}=\frac{\sqrt{1-\frac{v^{2}}{c^{2}}}}{v_{\mathsf{ph}}v}%
\hbar\omega
\end{equation}
and hence, the energy is
\begin{equation}
E=\frac{c^{2}}{v_{\mathsf{ph}}v}\hbar\omega.
\end{equation}

The above formula for the energy can be considered as a generalization of
Einstein-Planck formula for \textquotedblleft a photon in a
medium\textquotedblright. The factor
\begin{equation}
\frac{c^{2}}{v_{\mathsf{ph}}v}=\frac{nc}{v}%
\end{equation}
is due to the interaction with matter and becomes $1$ in the case of free
photons or other free particles.

\begin{remark}
In terms of a modified angular frequency $\tilde{\omega}=\frac{c^{2}%
}{v_{\mathsf{ph}}v}\omega$, this energy can be represented as $E=\hbar
\tilde{\omega}$, just similarly to the matter wave frequency for a free
massive particle. It is, however, sufficient for us to make use of $\omega$
(classical wave frequency) in the present paper.
\end{remark}

On the other hand, if we adopt the Abraham type formula
\begin{equation}
p=p_{\mathrm{free}}\cdot\frac{v}{c}=\frac{\hbar\omega v}{c^{2}}=\frac
{v_{\mathsf{ph}}v}{c^{2}}\hbar k
\end{equation}
then $m_{\mathsf{eff}}$ and $E$ should be given by
\begin{equation}
m_{\mathsf{eff}}=\frac{\sqrt{1-\frac{v^{2}}{c^{2}}}}{c^{2}}\hbar\omega
\end{equation}
and
\begin{equation}
E=\hbar\omega.
\end{equation}
This is identical to usual Einstein-Planck formula, where the Abraham momentum
itself violates the usual de Broglie formula.

To sum up, Minkowski picture provides
\begin{align}
\text{$\mathbf{p}$}=  &  \hbar\text{$\mathbf{k.}$}\\
E=  &  \frac{c^{2}}{v_{\mathsf{ph}}v}\hbar\omega\nonumber
\end{align}
while Abraham picture provides
\begin{align}
\text{$\mathbf{p}$}=  &  \frac{v_{\mathsf{ph}}v}{c^{2}}\hbar\text{$\mathbf{k.}%
$}\\
E=  &  \hbar\omega\nonumber
\end{align}

As we have seen above, the precise form of the effective mass of a photon in a
medium is in fact related to the question which generalized Einstein-de
Broglie relation for the case with interaction is suitable. This also suggests
the role of our effective mass as an order parameter due to certain kind of
symmetry breaking.

Perhaps (in the case of $n>1$) the Abraham type momentum, energy and mass can
be considered as those of \textquotedblleft pure photon
components\textquotedblright\ in a polariton, whose momentum (sum of momenta
for all branches) is showed to be Minkowski type \cite{BAR}. Further
discussion will be done in a succeeding paper \cite{OS}.

\end{document}